\documentclass[conference]{IEEEtran}
\IEEEoverridecommandlockouts
\usepackage{cite}
\usepackage{amsmath,amssymb,amsfonts}
\usepackage{graphicx}
\usepackage{textcomp}
\usepackage{xcolor}
\usepackage{fancyhdr}
\usepackage[hyphens]{url}
\usepackage{multirow}
\usepackage{algorithm}
\usepackage{lipsum}
\usepackage[noend]{algorithmic}

\usepackage{array}

\usepackage{array}

\newcolumntype{C}[1]{>{\centering\arraybackslash}m{#1}}
\newcolumntype{L}[1]{>{\raggedright\arraybackslash}m{#1}}
\newcolumntype{N}{@{}m{0pt}@{}}

\newcolumntype{P}[1]{>{\centering\arraybackslash}p{#1}}
\newcolumntype{M}[1]{>{\centering\arraybackslash}m{#1}}

\usepackage[pass]{geometry}
\usepackage[normalem]{ulem}
\usepackage{hyperref}
\usepackage{soul}

\usepackage{multicol}

\usepackage{tikz}
\newcommand*\circled[1]{\tikz[baseline=(char.base)]{
            \node[shape=circle,draw,inner sep=2pt] (char) {#1};}}

\def\BibTeX{{\rm B\kern-.05em{\sc i\kern-.025em b}\kern-.08em
    T\kern-.1667em\lower.7ex\hbox{E}\kern-.125emX}}
\begin{document}

\definecolor{myyellow}{RGB}{191, 144, 0}
\definecolor{myblue}{RGB}{60, 120, 216}
\definecolor{mypurple}{RGB}{103, 78, 167}
\definecolor{mygreen}{RGB}{106, 168, 79}
\definecolor{mypink}{RGB}{244, 204, 204}

\title{CoDR: Computation and Data Reuse Aware \\ CNN Accelerator
}

\author{
\IEEEauthorblockN{Alireza Khadem, Haojie Ye, and Trevor Mudge}
\IEEEauthorblockA{\textit{Computer Science and Engineering} \\
\textit{University of Michigan}\\
\{arkhadem, yehaojie, tnm\}@umich.edu}
}
\maketitle

\begin{abstract}

\underline{Co}mputation and \underline{D}ata \underline{R}euse is critical for the resource-limited Convolutional Neural Network (CNN) accelerators. This paper presents \textit{Universal Computation Reuse} to exploit weight sparsity, repetition, and similarity simultaneously in a convolutional layer. Moreover, \textit{CoDR} decreases the cost of weight memory access by proposing a customized Run-Length Encoding scheme and the number of memory accesses to the intermediate results by introducing an input and output stationary dataflow. Compared to two recent compressed CNN accelerators \cite{hegde2018ucnn}\cite{parashar2017scnn} with the same area of 2.85 mm$^2$, \textit{CoDR} decreases SRAM access by 5.08$\times$ and 7.99$\times$, and consumes 3.76$\times$ and 6.84$\times$ less energy.

\end{abstract}

\begin{IEEEkeywords}
CNN accelaration, computation reuse, data reuse, Run-Length Encoding
\end{IEEEkeywords}

\section{Introduction}

With the increasing complexity of the neural networks, the network model becomes even "deeper," improving the accuracy progressively at the cost of more computation resources and memory space, which requires high-performance processors and high-bandwidth memory systems. As a result, various Convolutional Neural Network (CNN) accelerators are proposed mainly to address two obstacles: (a) convolutional computation throughput, and (b) on-chip data usage efficiency. 

\begin{figure}
\centering
\includegraphics[width=1.0\columnwidth]{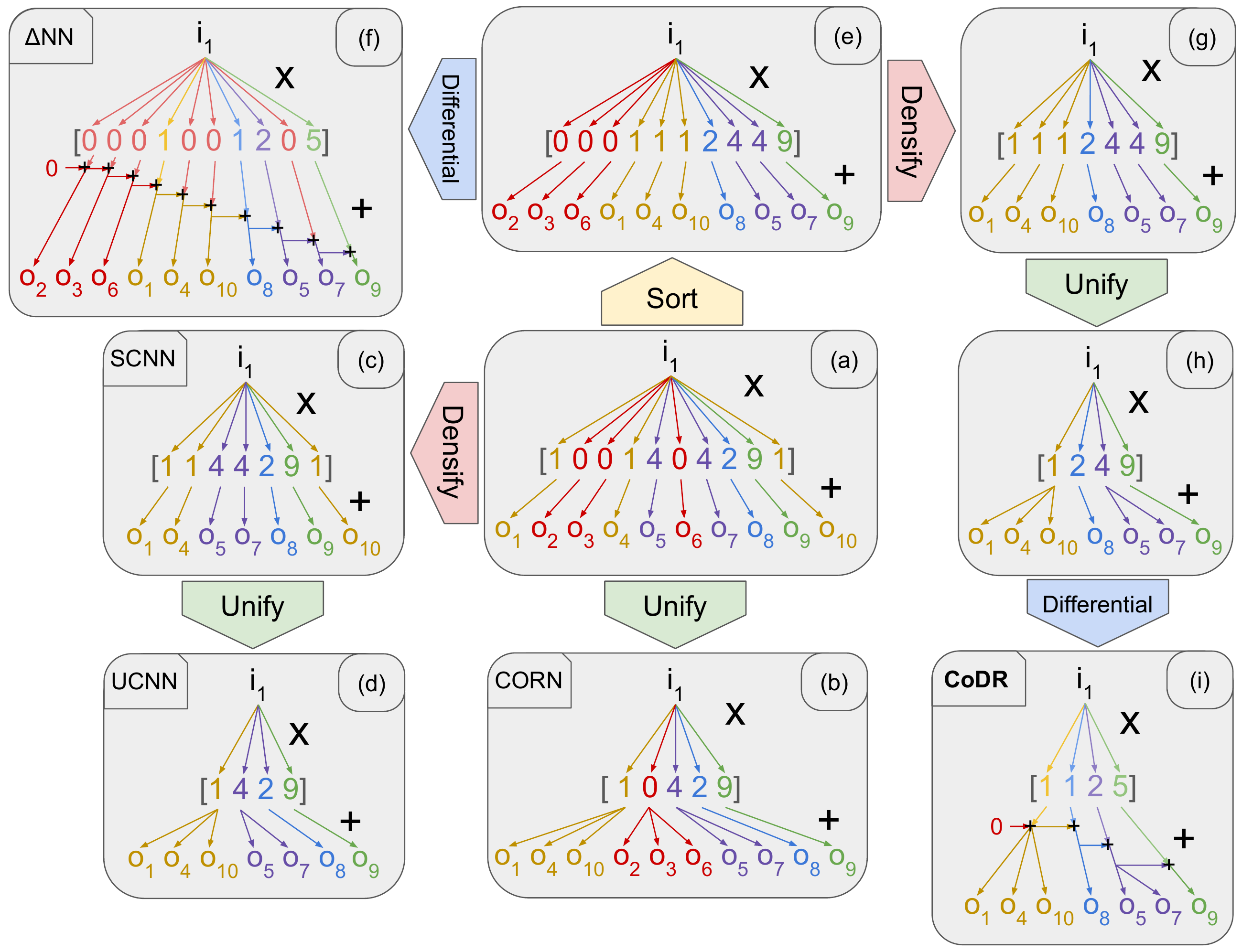}
\vspace*{-5mm} \caption{Dataflow comparison of \textit{SCNN} \cite{parashar2017scnn}, \textit{UCNN} \cite{hegde2018ucnn}, \textit{CORN} \cite{mahdiani2020computation}, \textit{$\Delta$NN} \cite{mahdiani2019delta}, and \textit{CoDR}. \textit{Densify} and \textit{Unify} eliminate ineffectual and repetitive weights, respectively. \textit{Differential} adds differential computation to exploit prior results.}
\label{fig:Comparison}
\vspace*{-5mm}
\end{figure}

To increase the computation throughput, specialized CNN dataflows exploit the characteristics of the network models, such as weight sparsity \cite{hegde2018ucnn, parashar2017scnn}, weight repetition \cite{hegde2018ucnn, mahdiani2020computation}, and weight similarity \cite{mahdiani2019delta}. Fig. \ref{fig:Comparison} illustrates these optimizations in the simple multiplication model of a fully-connected layer. To achieve high efficiency in data transfer (off-chip memory access), previous works \cite{han2016eie, parashar2017scnn, hegde2018ucnn} show that dataflow of CNNs can be pruned, quantized or compressed. Moreover, CNN accelerators exploit the locality in the off-chip data access by caching the recently-used data. To utilize the on-chip data, various CNN dataflows are proposed: \textit{Input Stationary} \cite{parashar2017scnn}, \textit{Output Stationary} \cite{du2015shidiannao, moons201714}, and \textit{Weight Stationary} \cite{jouppi2017datacenter, chi2016prime}, in which input features, output features, and weights are kept stationary in the processing element.

In this paper, we study three complementary computation reuse techniques proposed in \cite{hegde2018ucnn, parashar2017scnn, mahdiani2019delta}. \textit{CoDR} presents a novel CNN dataflow that employs scalar-matrix multiplication (Fig. \ref{fig:CNNPlan}b) to pave the way for the \textit{Universal Computation Reuse} (Fig. \ref{fig:Comparison}i) that explits weight sparsity, repetition, and similarity simultaneously in the convolutional layers. Next, we design a novel \textit{Run-Length Encoding} (RLE) scheme customized for the data values required data for the \textit{Universal Computation Reuse}. Finally, we observe that specialized CNN dataflows must make use of the accelerator characteristics; since the weights are compressed, access to the on-chip weights is less costly than the access to the input and output features. Thus, we design the loop ordering of the \textit{CoDR} dataflow to reduce the number of costly accesses to the input and output features. 

Compared to \textit{Sparse CNN} (SCNN) \cite{parashar2017scnn} and \textit{Unique Weight CNN} (UCNN) \cite{hegde2018ucnn} accelerators, \textit{CoDR}: (a) improves weight compression by $1.69\times$ and $2.80\times$ due to the RLE customization, (b) reduces SRAM accesses by $7.99\times$ and $5.08\times$ because of the dataflow loop ordering, and (c) achieves energy savings by $6.84\times$ and $3.76\times$ thanks to the \textit{universal computation reuse}. The contributions of this paper are as follows:
\begin{itemize}

\item{We introduce \textit{Universal Computation Reuse} that exploits weight sparsity, repetition, and similarity by adapting scalar-matrix multiplication for the convolutional layers.} 

\item{We customize RLE scheme to encode each type of weight data required for the \textit{Universal Computation Reuse}.}

\item{We present a CNN dataflow optimized for minimum on-chip memory accesses. \textit{CoDR} dataflow leverages the low per-access \textbf{cost} of weight memory access to reduce the \textbf{total number} of accesses to the input and output features by keeping them stationary in the processing elements.}

\item{We implement \textit{CoDR} architecture using a 45nm technology and compare it with two similar works: \textit{SCNN} \cite{parashar2017scnn} and \textit{UCNN} \cite{hegde2018ucnn}, which exploit weight sparsity and repetition.}
\end{itemize}

The rest of the paper is organized as follows: Section \ref{CompReuseSec} describes three complementary computation reuse optimizations. Section \ref{DataReuseSec} explains data reuse techniques in the \textit{CoDR} dataflow. Section \ref{ArchitectureSec} presents \textit{CoDR} architecture and Section \ref{EvaluationSec} compares \textit{CoDR} with two compressed CNN accelerators.

\section{Computation Reuse} \label{CompReuseSec}

\begin{figure}[t]
\centering
\includegraphics[width=0.5\textwidth]{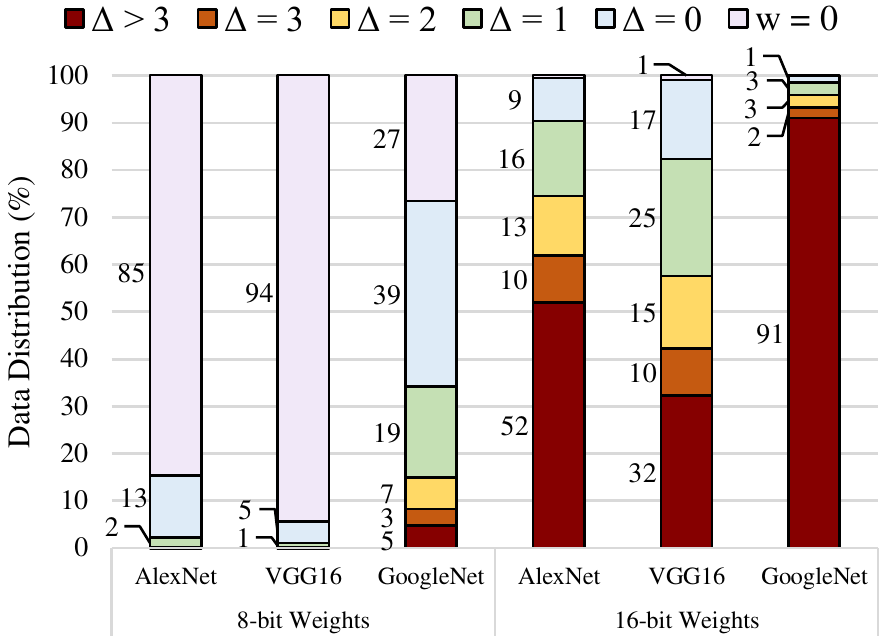}
\vspace*{-5mm}
\caption{Average distribution of the 8-bit and 16-bit zero weights and weight $\Delta$s (difference between sorted weights). Values less than 1\% are not shown.}
\vspace*{-4mm}
\label{fig:Data_Distribution}
\end{figure}

Fig. \ref{fig:Comparison} explains three complementary techniques of computation reuse used in the state-of-the-art neural network accelerators and introduces \textit{universal computation reuse} that employs all three methods simultaneously. These techniques are applied on the multiplication model of a fully-connected layer (Fig. \ref{fig:Comparison}a) in which an input feature ($i_1$) is multiplied by a vector of weights whose results are routed to the output accumulators. Since these modifications lead to irregularity of computation,
CNN accelerators use extra data to store output indexes. Section \ref{DataReuseSec} explains how CoDR maps the dataflow of the convolutional layers to this multiplication model.


\subsection{Weight Sparsity} \label{WeightSparsitySubSec}

Fig. \ref{fig:Data_Distribution} shows that the sparsity (W=0) of the 8-bit weight values of three contemporary CNN models can reach to $94\%$ in the VGG16 model \cite{simonyan2014vgg16}. Weight sparsity leads to ineffectual computation (red color in Fig. \ref{fig:Comparison}) that can be eliminated by \textbf{Densification}. Fig. \ref{fig:Comparison}c shows that \textit{SCNN} \cite{parashar2017scnn} exploits weight sparsity by removing zero terms from the on-chip data. 

\subsection{Weight Repetition} \label{WeightRepetitionSubSec}

While deep neural network inference requires millions of the weights (130 million weights in VGG16 \cite{simonyan2014vgg16}), number of unique weights is bounded by the data bit-length (256 unique weights for the 8-bit fixed-point numbers). This results in the computation redundancy (unique colors in Fig. \ref{fig:Comparison}). Fig. \ref{fig:Data_Distribution} shows that redundant computation on the non-zero weights ($\Delta$=0) can even reach to $39\%$ in the 8-bit GoogleNet \cite{szegedy2015googlenet} weights. Consequently, various CNN accelerators exploit the computation redundancy by the means of \textbf{Unification}; instead of multiplying all weights by the input feature, Fig. \ref{fig:Comparison}b shows that \textit{CORN} \cite{mahdiani2020computation} routes redundant computations through a programmable crossbar to the next-stage accumulation buffer to bypass the repeated computation. Fig. \ref{fig:Comparison}d presents \textit{UCNN} \cite{hegde2018ucnn} that factorizes out the same weights in the dot product as an \textit{activation group}, which includes the coordinates of the input features that are multiplied by the same weight. \textit{UCNN} also exploits weight sparsity by eliminating the activation groups related to the zero weights. As a result of \textit{Unification}, number of multiplications is reduced to the number of unique weights.

\subsection{Differential Computation} \label{DifferentialComputationSubSec}

Fig. \ref{fig:Data_Distribution} shows that the portion of zero and redundant weights drop significantly to $0.5\%$ and $9.0\%$ in the 16-bit fixed-point weights, which makes the \textit{Densification} and \textit{Unification} techniques useless. As an alternative, \textbf{Differential} computation operates on the differences of the similar weights rather than the absolute operands by reusing the previous results. Fig. \ref{fig:Comparison}f shows that \textit{$\Delta$NN} \cite{mahdiani2019delta} computes the difference between successive sorted weights ($\Delta$), and multiplies the $\Delta$s instead of the absolute weights. Equation (\ref{Delta_eq}) shows how this accelerator exploits differential computation in the CNN inference.
\begin{equation} \label{Delta_eq}
\begin{split}
    w_{m+1} \times i &= (w_{m+1} - w_{m}) \times i +  w_{m} \times i \\
    &=\Delta_{m} \times i +  w_{m} \times i
\end{split}
\end{equation}

Fig. \ref{fig:Data_Distribution} shows that using \textit{Differential Computation} enables the CNN accelerator to extend the computation reuse to the small $\Delta$ values and compensates the lack of weight sparsity and weight repetition in the 16-bit data. However, \textit{$\Delta$NN} \cite{mahdiani2019delta} is unaware of weight sparsity (W=0) and repetition ($\Delta$=0).

\subsection{Universal Computation Reuse}

While computation reuse techniques are employed individually in different works, we observed that they are complementary to each other. Thus, we introduce \textit{universal computation reuse} that employs \textit{Densification}, \textit{Unification}, and \textit{Differential Computation} simultaneously to exploit W=0, $\Delta$=0, and small $\Delta$ values. The following steps that implement \textit{universal computation reuse} impose no overhead on the chip as they are executed offline once per each neural network model.

(i) We break a convolutional layer into the tiles of $T_N$ input and $T_M$ output channels that are processed at the same time by a processing unit of the \textit{CoDR} accelerator. (ii) Tiles of weight and bias terms are quantized into the 8-bit fixed-point numbers. (iii) We collect the weights related to a single input channel inside the tiles, and produce \textbf{$T_N$ weight vectors}, each contains the weights of $T_M$ weight kernel for a unique input channel. (iv) We sort, densify, and unify the weights (Fig. \ref{fig:Comparison}e, g, and h). (v) $\Delta$ values of the non-zero unique weights and their corresponding indexes are computed and sent to the RLE encoders (Section \ref{RLE_SubSec}) for the data reuse techniques.



\section{Data Reuse} \label{DataReuseSec}

\begin{figure}[t]
\centering
\includegraphics[width=0.48\textwidth]{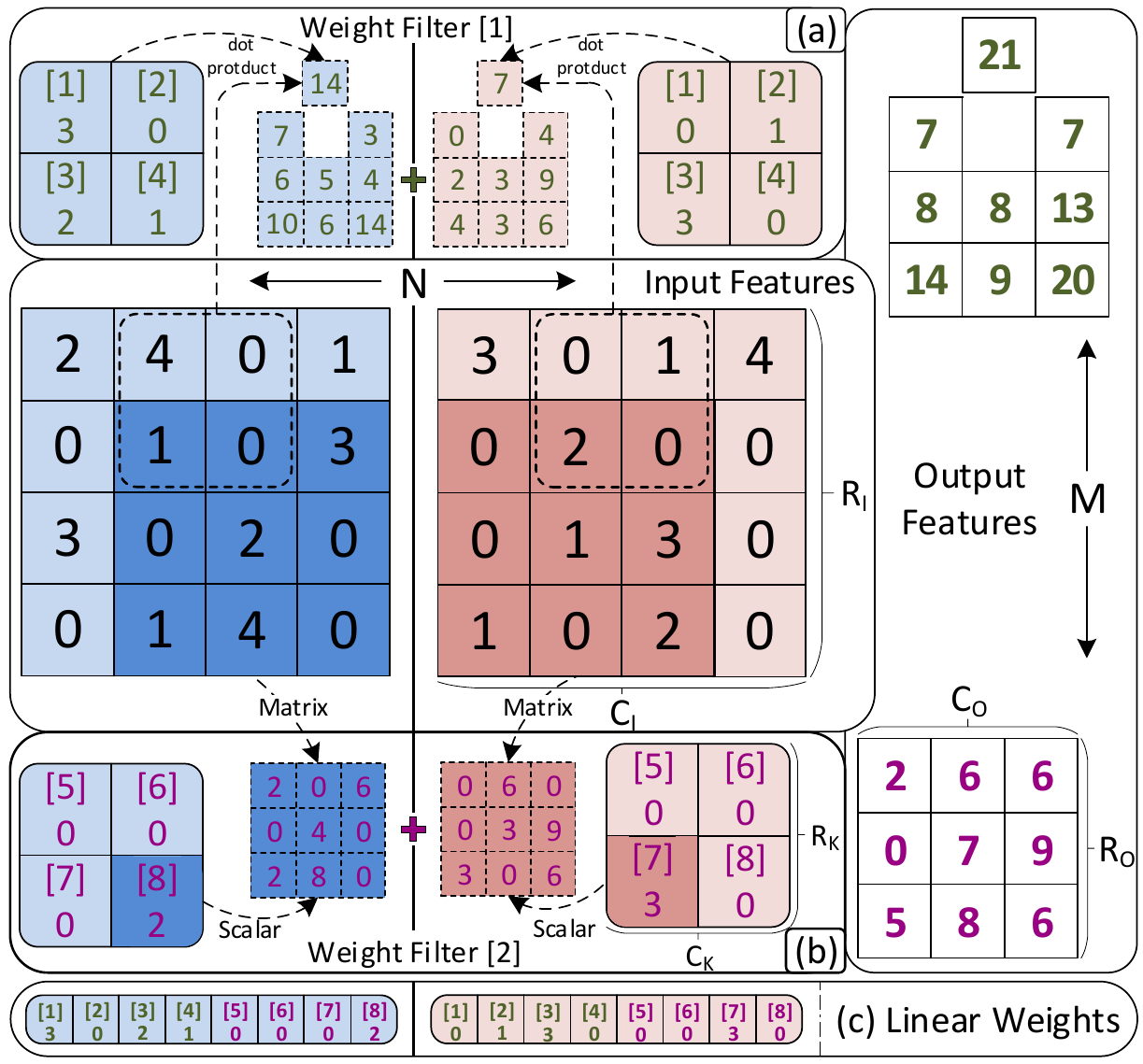}
\vspace*{-2mm}
\caption{A convolutional layer with $N$=2 input channels (blue and red cells), $M$=2 output channels (green and purple numbers), input feature size of $R_I$=$C_I$=4, kernel size of $R_K$=$C_K$=2 and output feature size of $R_O$=$C_O$=3. (a) Conventional CNN inference uses 3D convolution operations illustrated for the the first output channel. (b) Instead, \textit{CoDR} employs scalar-matrix multiplication for CNN inference as shown for the second output channel. (c) Linearized weight vectors of the first (blue) and second (red) input channels.}
\vspace*{-4mm}
\label{fig:CNNPlan}
\end{figure}

Data transfer in neural network accelerators costs more than computation. According to our evaluations, \textit{UCNN}\cite{hegde2018ucnn} and \textit{SCNN}\cite{parashar2017scnn} spend $64\%$ and $76\%$ of the total energy for data transfer. We show the \textit{CoDR} data reuse scheme below. 

\subsection{Scalar-Matrix Multiplication in CoDR Dataflow}

Convolutional layers have a 4-dimensional (4D) weight that consists of 3D \textit{weight filters} for $M$ output channels, each of which contains $R_K \times C_K$ \textit{weight kernels} for $N$ input channels. Output features can be produced by applying an activation function on each of the following operations results:

\textbf{3D convolutions}. Fig. \ref{fig:CNNPlan}a shows how the features of the first output channel are computed by a 3D convolution of the weight filter and the input features. A single output feature is computed by 2D dot product operations, each of which includes pairwise multiplication of a weight kernel by a window of the input features, followed by accumulating the partial products. 2D dot product results ($14$ and $7$) are then accumulated across input channels to produce the result of the 3D convolution ($21$). The input feature window moves by $stride$ cells to compute the next output feature.

\textbf{Scalar-matrix Multiplication.} Another CNN dataflow proposed by \cite{lu2019efficient} is shown in Fig. \ref{fig:CNNPlan}b. Shaded regions of the input features present the cells involved in the dot product operations between the non-zero weights of the second weight filter ($2$ and $3$) and the input features. Each weight (scalar) is multiplied by its corresponding region of the input features (matrix). Then, partial results of a weight filter (dotted matrices) are accumulated to produce the final results.

\textit{CoDR} dataflow employs the scalar-matrix multiplication model since it breaks the dependency between the individual weight terms and enables us to linearize the weight kernels (Fig. \ref{fig:CNNPlan}c) to make use of the correlation between weight terms. Comparing with Fig. \ref{fig:Comparison}, each input feature in the shaded region is mapped to a single input ($i_1$) and weight vectors are manipulated by sorting, densifying, unifying, and $\Delta$ calculation to exploit weight sparsity, repetition, and similarity.

\subsection{On-Chip Data Reuse (Dataflow Loop Ordering)}

We observe that a CNN dataflow must emphasize on the features in which the accelerator outperforms. As \textit{CoDR} employs novel RLE schemes to compress the weights (Section \ref{RLE_SubSec}), accesses to the weight terms (on average $1.69$ bits/weight) are less costly than access to the input or output features ($8$ bit/feature). \textit{CoDR} reduces the number of on-chip accesses to the input and output features by using an input and output stationary dataflow whose loop ordering is illustrated in Fig. \ref{fig:ProcessingUnit}a. \circled{4}, \circled{3}, and \circled{2} show that \textit{CoDR} accesses output features only once and \circled{3}, \circled{2}, and \circled{1} show that input features are fetched $\frac{M}{T_{PU}\times T_M}$ times in which $T_{PU}$ is the number of processing units. In contrast, \textit{UCNN} \cite{hegde2018ucnn} and \textit{SCNN} \cite{parashar2017scnn} increase the number of costly accesses to the input and output features.

\subsection{Run-Length Encoding} \label{RLE_SubSec}


\begin{figure}[t]
\centering
\includegraphics[width=0.49\textwidth]{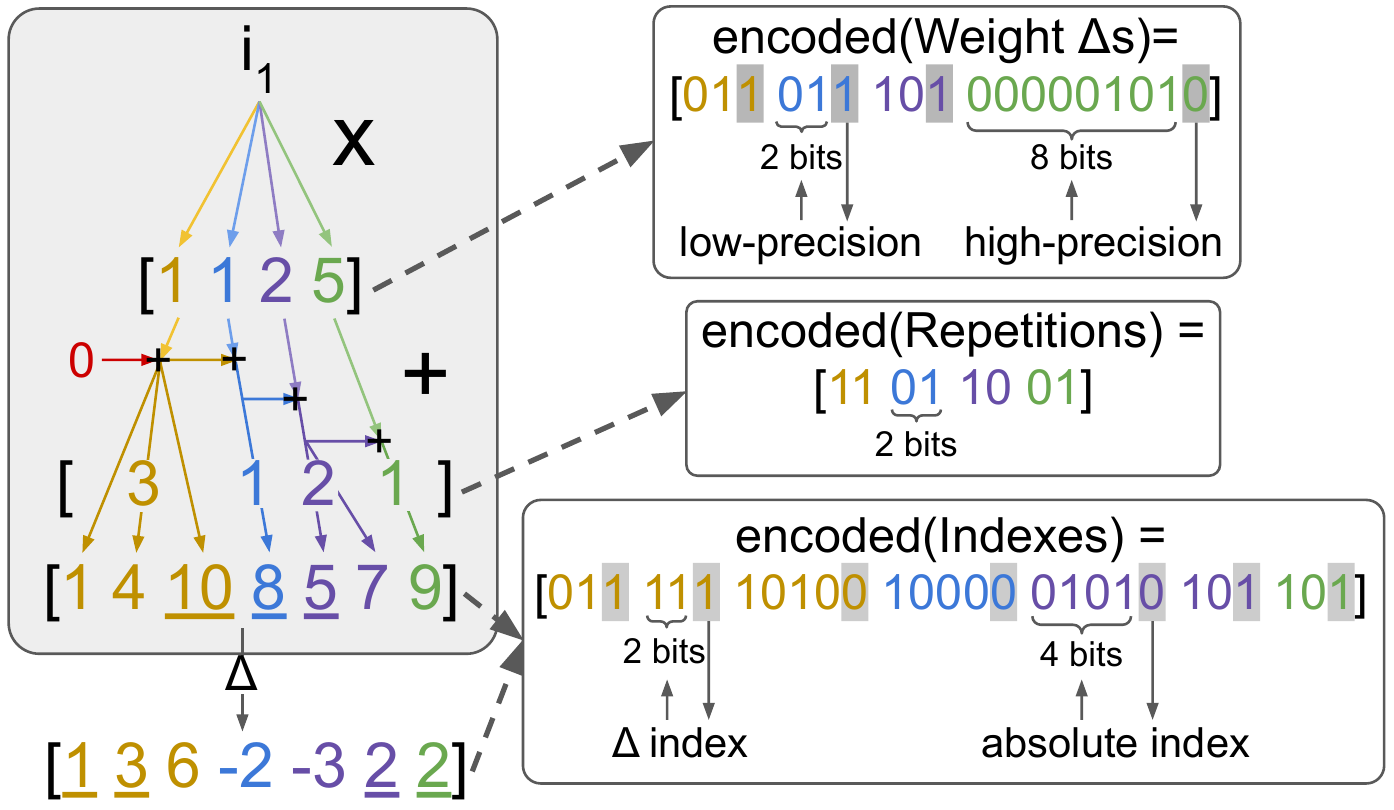}
\vspace*{-6mm}
\caption{Customized run-length encoding of the Fig. \ref{fig:Comparison}i example. Non-zero unique weight $\Delta$s, unique weight repetitions, and indexes are all encoded with the encoding parameter of 2. Least significant bits of the compressed weight $\Delta$s and indexes show if data is encoded with the high/low-precision and absolute/$\Delta$ mode, respectively.}
\vspace*{-5mm}
\label{fig:Encoding}
\end{figure}

Due to the irregularity of computation arising from employing the \textit{universal computation reuse} (Fig. \ref{fig:Comparison}i), \textit{CoDR} stores three data structures to obtain the computation order: (a) $\Delta$ values between the non-zero unique weights. (b) Output indexes related to each unique weight repetition. (c) And the count of unique weight repetitions. \textit{CoDR} customizes RLE schemes for the characteristics of each data type. Additionally, the RLE process of each data structures is independent; \textit{RLE Encoder} iterates on the encoding parameter of each data structure, finds the specific parameters with which it encodes them in the least memory space, and stores the encoding parameters along with the compressed values to the off-chip memory. The per-structure and per-layer customization improves the compression rate and DRAM access efficiency. Fig. \ref{fig:Encoding} shows the encoding process of the example in Fig. \ref{fig:Comparison}i.

\textbf{Unique Weight $\Delta$} values depend on the weight repetition, and thus on the precision of the raw weights. For example, Fig. \ref{fig:Data_Distribution} shows that $\Delta$ values are smaller in the 8-bit raw data since the resolution is less than the 16-bit weight terms and weights are highly repeated. \textit{RLE Encoder} exploits the small $\Delta$ values by encoding them in narrower numbers, e.g., 2 bits. When a $\Delta$ does not fit into this bit-length, it is encoded as a full-precision value, i.e., 8 bits. A single bit (shaded bits in Fig. \ref{fig:Encoding}) is appended to each $\Delta$ value to distinguish between the low-precision and high-precision values. A longer low-precision bit-length results in more weights encoded in the low-precision values, yet, each requires more memory space.

\textbf{Unique Weight Repetition.} The possible repetition count for unique weights ranges from $1$ to $T_{M} \times R_{K} \times C_{K}$. \textit{RLE Encoder} encodes them in numbers with a specific bit-length. When the repetition count of a unique weight overflows, a dummy unique weight with $\Delta = 0$ is inserted to the unique weight data structure to track the overflowed portion of the repetition count. A long bit-length decreases the overflows, but it wastes the memory space for the small repetitions.

\textbf{Indexes.} \textit{CoDR} employs the same RLE scheme as the unique weight $\Delta$s with one difference: Having computed the $\Delta$ values between subsequent indexes, \textit{RLE Encoder} uses absolute indexes when the $\Delta$ value is either negative or does not fit into the low-precision bit-length.


\section{CoDR Architecture} \label{DeltaAccSec}  \label{ArchitectureSec}

\begin{figure}[t]
\centering
\includegraphics[width=0.49\textwidth]{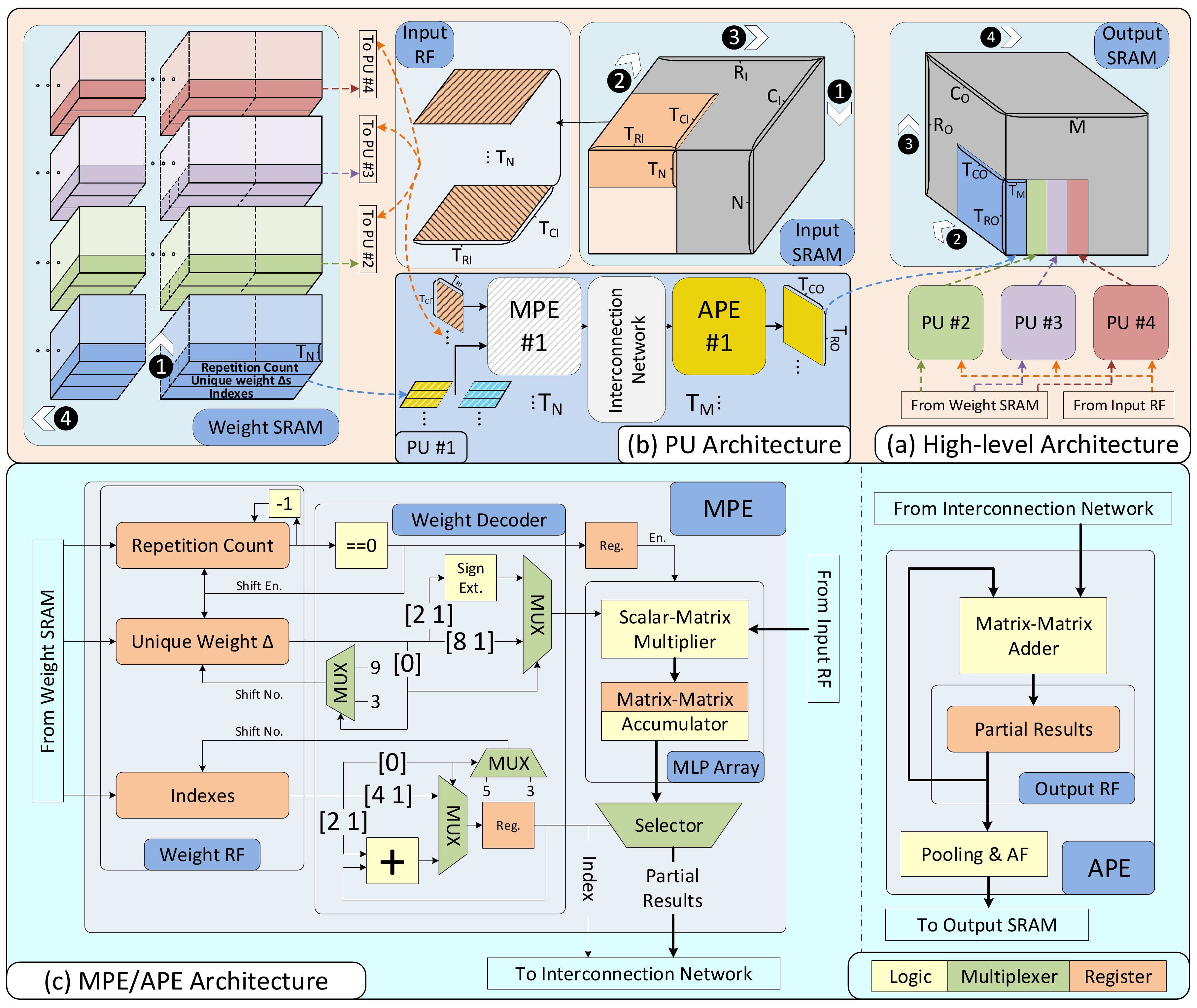}
\vspace*{-5mm}
\caption{(a) High-level \textit{CoDR} architecture includes Weight, Input, and Output SRAM modules, Input Register File (RF), and Processing Units (PU). Circled numbers show the loop ordering of the \textit{CoDR} dataflow. (b) PU architecture consists of $T_N$ Multiplier Processing Elements (MPE), $T_M$ Accumulator Processing Units (APE), and an Interconnection Network. (c) MPE decodes compressed weight structures, multiplies scalar weights by the input feature matrices, selects the results based on the index, and sends them to the corresponding APE, which accumulates and post-processes the partial results.}
\vspace*{-4mm}
\label{fig:ProcessingUnit}
\end{figure}

\subsection{High-level Architecture}

Figure \ref{fig:ProcessingUnit}a illustrates \textit{CoDR} architecture that contains $3$ SRAM cells for input features, weights, output features, and $T_{PU}$ (=4) Processing Units (PUs). Since all PUs work on the same region of the input/output features, an input register file (RF) shared between all PUs caches input features. A PU calculates $T_M$ tiles of $T_{RO}\times T_{CO}$ output features in an \textit{Iteration}, which is executed in multiple \textit{Cycles}. In each \textit{Cycle}, $T_N$ tiles of $T_{RI} \times T_{CI}$ input features are processed.

\subsection{Processing Unit Architecture}

A processing unit contains $T_N$ Multiplier Processing Elements (MPE), $T_M$ Accumulator Processing Elements (APE), and an interconnection network that connects the MPEs to APEs (Fig. \ref{fig:ProcessingUnit}b). In each \textit{Iteration}, APE accumulates the partial results of a single output channel; a MPE is assigned with an input channel in a \textit{Cycle}; the $i^{th}$ layer of the input features inside the Input RF is broadcasted to the $i^{th}$ MPE of all PUs. 

\subsection{MPE and APE Architecture}
Fig. \ref{fig:ProcessingUnit}c shows that a Weight RF gradually loads the compressed Repetition Counts, Unique Weight $\Delta$s, and Indexes from the Weight SRAM. These data structures are then decoded by the Weight Decoder module. A differential Scalar-Matrix Multiplier (MLP Array) multiplies each unique weight (scalar) by a tile of input features from the Input RF (matrix). Based on the weight coordinates, the interconnection network sends the selected partial products to the destination APE, where the partial results are added with the accumulated results. At the end of an \textit{Iteration}, final results are fed to the pooling layer and activation function logic for post processing.

\subsection{Computation and Data Reuse Support} \label{CoDRSupportToAccSec}

\textit{Universal computation reuse} is employed in the \textit{CoDR} architecture: (i) By eliminating zero weights from the off-chip data, \textit{CoDR} implicitly exploits weight sparsity. (ii) To make use of weight repetition, Weight RF broadcasts a unique weight to the MLP Array where it is multiplied by all of the input features. In the subsequent clock cycles, an index related to each repetition of the unique weight is fetched for the \textit{Selector} to choose a $T_{RO} \times T_{CO}$ window of the multiplication results and send them to the corresponding APE. (iii) Differential computation is employed by adding a Matrix-Matrix Accumulator inside the MLP Array to add the $\Delta$ multiplication results to the prior accumulated results. \textit{Data reuse} is also used by (i) keeping the input features stationary in the Input RF (MPE) and output features in the Output RF (APE). (ii) Loop ordering (circled numbers) accesses output features only once (fully output stationary) and input features only $\frac{M}{T_M\times T_{PU}}$ times (semi input stationary). (iii) Weights are all compressed and decoded by the Weight Decoder.

\begin{table}
\caption{\label{tab:DeltaAccConfig}RTL design tiling parameters.}
\begin{center}
\begin{tabular}{ | M{2cm} | M{1.2cm} | M{1.2cm} | M{1.2cm} | } 
\hline
Parameter & CoDR & UCNN & SCNN \\
\hline\hline
$T_{PU}$ & 8 & 48 & 21 \\
\hline
$T_M$, $T_N$ & 4, 4 & 1, 4 & 2, 1 \\
\hline
$T_{RO}$, $T_{CO}$ & 8, 8 & 1, 8 & 1, 1 \\ 
\hline
$T_{RI}$, $T_{CI}$ & 20, 20 & 1, 12 & 1, 1 \\ 
\hline
$\times$ per PU & 64 & 8 & 16 \\
\hline
\end{tabular}
\vspace*{-5mm}
\end{center}
\end{table}

\begin{figure}[t]
\centering
\hspace*{-0.1in}
\includegraphics[width=0.49\textwidth]{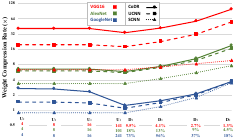}
\caption{Weight compression analysis for three CNN models with different weight densities and repetitions. The middle group shows the original density and repetition while they drop towards the right-side and left-side groups.}
\vspace*{-3mm}
\label{fig:Compression}
\end{figure}

\begin{figure}[t]
\centering
\hspace*{-1mm}
\includegraphics[width=0.49\textwidth]{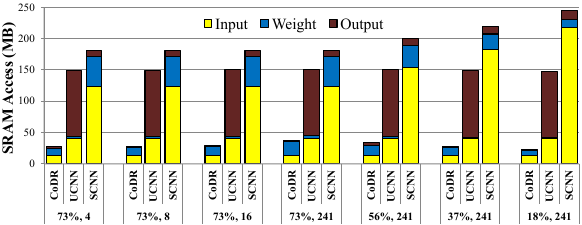}
\caption{SRAM Access analysis of the GoogleNet model with different weight densities and repetitions. Input and output stationary dataflow and customized RLE scheme reduce SRAM access compared to \textit{UCNN} \cite{hegde2018ucnn} and \textit{SCNN} \cite{parashar2017scnn}.}
\vspace*{-3mm}
\label{fig:SRAM}
\end{figure}

\section{Evaluation} \label{EvaluationSec}

\subsection{Benchmarks and Analysis Tools}

We compare \textit{CoDR} with two recent compressed CNN accelerators: \textit{SCNN}\cite{parashar2017scnn} and \textit{UCNN}\cite{hegde2018ucnn} that exploit weight sparsity and repetition, respectively. We implement all designs in Verilog and synthesize them using Synopsys Design Compiler with a 45 nm technology. Table \ref{tab:DeltaAccConfig} presents the configuration of the three architectures. We assign the number of processing units ($T_{PU}$) to equalize the overall area of each design. We employ $250$ kB of input and output SRAM cells and $200$ kB of weight SRAM such that they accommodate all data required for an \textit{Iteration}. Consequently, overall area of each architecture is 2.85 mm$^2$. We evaluate the number of memory accesses in each design by cycle-accurate simulation. SRAM cells are modeled using CACTI \cite{muralimanoharXcacti} and DRAM access energy consumption is considered $160$ pJ/B \cite{hegde2018ucnn}. We quantize AlexNet \cite{krizhevsky2012alexnet}, VGG16 \cite{simonyan2014vgg16}, and GoogleNet \cite{szegedy2015googlenet} weights into the 8-bit fixed-point terms. We evaluate four weight densities (denoted as \textit{D}) by randomly eliminating the non-zero weights and study different numbers of unique weights (denoted as \textit{U}) by making the $8-log_2(U)$ least significant bits of weights zero.

\begin{figure*}[t]
\centering
\hspace*{-0.13in}
\includegraphics[width=1\textwidth]{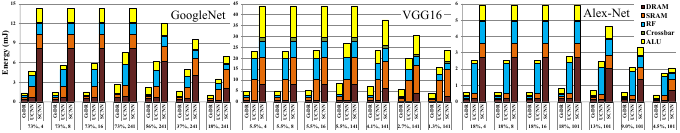}
\caption{\textit{CoDR} consumes on average $2.4\times$ and $4.8\times$ less energy compared to \textit{UCNN} \cite{hegde2018ucnn} and \textit{SCNN} \cite{parashar2017scnn}, respectively. Energy consumption of all accelerators decrease with weight density degradation (right-side groups) and also by limiting unique weights in \textit{CoDR} and \textit{UCNN} (left-side groups).}
\vspace*{-2mm}
\label{fig:Energy}
\end{figure*}

\subsection{Weight Compression Analysis} \label{WeightCompressionAnalysis}

Fig. \ref{fig:Compression} shows the compression rate across three CNN models with different weight densities and repetitions. \textit{SCNN} does not compress the non-zero weights and stores the number of zero values between two subsequent non-zero weights in $4$ bits \cite{parashar2017scnn}. \textit{UCNN} employs RLE to compress the weights and indexes \cite{hegde2018ucnn}, yet, it uses bit-length of $5$ for all layers. \textit{CoDR} chooses the optimal parameter with which it encodes weights and indexes into the minimal memory space. \textit{UCNN} additionally appends 1 bit to each index to indicate the transition to a new unique weight \cite{hegde2018ucnn}. \textit{CoDR} prevents this 1-bit overhead by encoding the repetition counts in numbers with a specific bit-length. This improves the compression for the highly-repeated weights. As a result, \textit{CoDR} compresses the weights by $1.69\times$ and $2.80\times$ more than \textit{UCNN} and \textit{SCNN}.

We observe that weight sparsity (right groups) and weight repetition (left groups) have different effects on the weight data values. Since all designs eliminate zero weights and their corresponding indexes form the off-chip memory, sparsity results in longer distance between subsequent non-zero weights and their indexes, and smaller number of unique weight repetitions. Thus, customized \textit{RLE encoder} chooses longer bit-lengths for weights and indexes, shorter bit-length for the repetition counts, and improves compression rate by $1.61\times$ and $4.75\times$ compared to \textit{UCNN} and \textit{SCNN}. On the other hand, limiting number of unique weights increases the weight repetition. Consequently, weight and index $\Delta$ values are smaller, and unique weights repeat more; customized \textit{RLE encoder} employs shorter bit-lengths for weights and indexes and longer bit-lengths for the repetition counts. Consequently, \textit{CoDR} compression rate is $1.87\times$ and $4.52\times$ higher than \textit{UCNN} and \textit{SCNN} in these groups.

\subsection{SRAM Access Analysis} \label{SRAMAnalysisSec}

Besides weight compression rate, dataflow substantially affects SRAM access as it determines the trade-offs in how PUs access each type of data. Since all designs compress weight terms, an access to the weight SRAM costs $20.61\times$ (\textit{CoDR}), $12.17\times$ (\textit{UCNN}), and $4.34\times$ (\textit{SCNN}) less than an access to the input or output features. As a result, a weight-compressed CNN dataflow should decrease the number of costly accesses to the input and output features by keeping them stationary in PEs. Thus, \textit{CoDR} dataflow increases the number of weight accesses to maximize input and output feature reuse ($50\%$ of the \textit{CoDR} SRAM bandwith is spent on the weights). On the other hand, \textit{UCNN} and \textit{SCNN} increase the number of input feature accesses by $20.4\times$ and $21.3\times$. Moreover, \textit{UCNN} accesses each output feature $72.1$ times, and spends only $1.40\%$ of the total SRAM bandwidth for the weight access. Having considered increased costly accesses to the input and output features by \textit{UCNN} and \textit{SCNN}, \textit{CoDR} dataflow reduces SRAM accesses by $5.08\times$ and $7.99\times$.

\subsection{Energy Consumption Analysis}

Fig. \ref{fig:Energy} illustrates the effect of weight density and repetition on the energy consumption. \textit{CoDR} consumes on average $3.76\times$ and $6.84\times$ less energy relative to \textit{UCNN} and \textit{SCNN}.

\textbf{DRAM, SRAM, and Register File Access.} Since the intermediate results are kept on-chip, input and output feature access consumes less than $15\%$ of total DRAM access energy in all designs. Thus, DRAM and SRAM energy footprints follow the same characteristics of Fig. \ref{fig:Compression} and \ref{fig:SRAM}. DRAM is the most energy-hungry part of the \textit{SCNN} design ($37\%$) due to the low compression rate while it consumes $18\%$ and $6\%$ of the \textit{CoDR} and \textit{UCNN} energy. \textit{UCNN} and \textit{SCNN} SRAM energy consumption does not change with weight sparsity and repetition, since $98.6\%$ and $86.4\%$ of total SRAM bandwidth is used by the input and output feature access. However, $50\%$ of the \textit{CoDR}'s SRAM energy is consumed by the weight access that drops by $16\%$ and $8\%$ in the right and left groups. Register file energy consumption includes the input, weight, and output RF access that is on average $342$, $3880$, and $2861$ $\mu J$ in \textit{CoDR}, \textit{UCNN}, and \textit{SCNN}, respectively.

\textbf{ALU and Crossbar.} ALU consumes a significant portion of \textit{CoDR} energy ($42\%$ on average) as DRAM and SRAM accesses are minimized with the RLE compression and \textit{CoDR} dataflow. ALU energy consumption of all designs decreases by around $25\%$ in the right groups with the density degradation. \textit{CoDR} and \textit{UCNN} also exploit weight repetition in the groups with the limited number of unique weights. Thus, ALU energy consumption drops substantially by $52\%$ and $48\%$ when limiting the number of unique weights to $16$. In short, ALU in \textit{CoDR} consumes $1.32\times$ and $3.80\times$ less energy than \textit{UCNN} and \textit{SCNN} due to the \textit{universal computation reuse}. Finally, crossbar is the least energy-hungry module in both \textit{CoDR} and \textit{UCNN} designs as it consumes $4.7\%$ and $2.3\%$ of total energy.

\section{Conclusion}

In this work, we study three complementary computation reuse optimizations for the CNN accelerators and introduce \textit{Universal Computation Reuse} that exploits weight sparsity, repetition, and similarity simultaneously. We propose a dataflow that employs scalar-matrix multiplication to apply \textit{Universal Computation Reuse} to the convolutional layers. \textit{CoDR} dataflow makes use of data reuse to minimize the on-chip memory access. We reduce the cost of each weight memory access by customizing run-length encoding based on the weight values. The loop ordering of the \textit{CoDR} dataflow also reduces the total number of accesses to the input and output features by keeping them stationary in the processing elements. Our evaluation over three CNNs with different weight densities and repetitions shows that compared to two recent compressed CNN accelerators with the equivalent area of $2.85$ mm$^2$, \textit{CoDR} requires $1.69\times$ and $2.80\times$ less DRAM access, reduces SRAM access by $5.08\times$ and $7.99\times$, and consumes $3.76\times$ and $6.84\times$ less energy.

\bibliographystyle{IEEEtranS}
\bibliography{refs}

\begin{thebibliography}{10}
\providecommand{\url}[1]{#1}
\csname url@samestyle\endcsname
\providecommand{\newblock}{\relax}
\providecommand{\bibinfo}[2]{#2}
\providecommand{\BIBentrySTDinterwordspacing}{\spaceskip=0pt\relax}
\providecommand{\BIBentryALTinterwordstretchfactor}{4}
\providecommand{\BIBentryALTinterwordspacing}{\spaceskip=\fontdimen2\font plus
\BIBentryALTinterwordstretchfactor\fontdimen3\font minus
  \fontdimen4\font\relax}
\providecommand{\BIBforeignlanguage}[2]{{%
\expandafter\ifx\csname l@#1\endcsname\relax
\typeout{** WARNING: IEEEtranS.bst: No hyphenation pattern has been}%
\typeout{** loaded for the language `#1'. Using the pattern for}%
\typeout{** the default language instead.}%
\else
\language=\csname l@#1\endcsname
\fi
#2}}
\providecommand{\BIBdecl}{\relax}
\BIBdecl

\bibitem{parashar2017scnn}
J.~Albericio, P.~Judd, T.~Hetherington, T.~Aamodt, N.~Enright~Jerger, and
  A.~Moshovo, ``Scnn: An accelerator for compressed-sparse convolutional neural
  networks,'' \emph{ACM SIGARCH Computer Architecture News}, vol.~45, no.~2,
  pp. 27--40, 2017.

\bibitem{chi2016prime}
P.~Chi, S.~Li, C.~Xu, T.~Zhang, J.~Zhao, Y.~Liu, Y.~Wang, and Y.~Xie, ``Prime:
  A novel processing-in-memory architecture for neural network computation in
  reram-based main memory,'' \emph{ACM SIGARCH Computer Architecture News},
  vol.~44, no.~3, pp. 27--39, 2016.

\bibitem{du2015shidiannao}
Z.~Du, R.~Fasthuber, T.~Chen, P.~Ienne, L.~Li, T.~Luo, X.~Feng, Y.~Chen, and
  O.~Temam, ``Shidiannao: Shifting vision processing closer to the sensor,'' in
  \emph{Proceedings of the 42nd Annual International Symposium on Computer
  Architecture}, 2015, pp. 92--104.

\bibitem{han2016eie}
S.~Han, X.~Liu, H.~Mao, J.~Pu, A.~Pedram, M.~A. Horowitz, and W.~J. Dally,
  ``Eie: efficient inference engine on compressed deep neural network,''
  \emph{ACM SIGARCH Computer Architecture News}.

\bibitem{hegde2018ucnn}
K.~Hegde, J.~Yu, R.~Agrawal, M.~Yan, M.~Pellauer, and C.~Fletcher, ``Ucnn:
  Exploiting computational reuse in deep neural networks via weight
  repetition,'' in \emph{2018 ACM/IEEE 45th Annual International Symposium on
  Computer Architecture (ISCA)}.\hskip 1em plus 0.5em minus 0.4em\relax IEEE,
  2018, pp. 674--687.

\bibitem{jouppi2017datacenter}
N.~P. Jouppi, C.~Young, N.~Patil, D.~Patterson, G.~Agrawal, R.~Bajwa, S.~Bates,
  S.~Bhatia, N.~Boden, A.~Borchers \emph{et~al.}, ``In-datacenter performance
  analysis of a tensor processing unit,'' in \emph{Proceedings of the 44th
  Annual International Symposium on Computer Architecture}.

\bibitem{krizhevsky2012alexnet}
\BIBentryALTinterwordspacing
A.~Krizhevsky, I.~Sutskever, and G.~E. Hinton, ``Imagenet classification with
  deep convolutional neural networks,'' in \emph{Advances in Neural Information
  Processing Systems 25}, F.~Pereira, C.~J.~C. Burges, L.~Bottou, and K.~Q.
  Weinberger, Eds.\hskip 1em plus 0.5em minus 0.4em\relax Curran Associates,
  Inc., 2012, pp. 1097--1105. [Online]. Available:
  \url{http://papers.nips.cc/paper/4824-imagenet-classification-with-deep-convolutional-neural-networks.pdf}
\BIBentrySTDinterwordspacing

\bibitem{lu2019efficient}
L.~Lu, J.~Xie, R.~Huang, J.~Zhang, W.~Lin, and Y.~Liang, ``An efficient
  hardware accelerator for sparse convolutional neural networks on fpgas,'' in
  \emph{2019 IEEE 27th Annual International Symposium on Field-Programmable
  Custom Computing Machines (FCCM)}.\hskip 1em plus 0.5em minus 0.4em\relax
  IEEE, 2019.

\bibitem{mahdiani2019delta}
H.~Mahdiani, A.~Khadem, A.~Ghanbari, M.~Modarressi, F.~Fattahi-Bayat, and
  M.~Daneshtalab, ``\uppercase{$\Delta$NN}: Power-efficient neural network
  acceleration using differential weights,'' \emph{IEEE Micro}, vol.~40, no.~1,
  pp. 67--74, 2019.

\bibitem{mahdiani2020computation}
H.~Mahdiani, A.~Khadem, A.~Yasoubi, A.~Ghanbari, M.~Modarressi, and
  M.~Daneshtalab, ``Computation reuse-aware accelerator for neural networks,''
  \emph{Hardware Architectures for Deep Learning}, p. 147, 2020.

\bibitem{moons201714}
B.~Moons, R.~Uytterhoeven, W.~Dehaene, and M.~Verhelst, ``14.5 envision: A
  0.26-to-10tops/w subword-parallel dynamic-voltage-accuracy-frequency-scalable
  convolutional neural network processor in 28nm fdsoi,'' in \emph{2017 IEEE
  International Solid-State Circuits Conference (ISSCC)}.\hskip 1em plus 0.5em
  minus 0.4em\relax IEEE, 2017, pp. 246--247.

\bibitem{muralimanoharXcacti}
N.~Muralimanohar and R.~Balasubramonian, ``Cacti 6.0: A tool to understand
  large caches.''

\bibitem{simonyan2014vgg16}
K.~Simonyan and A.~Zisserman, ``Very deep convolutional networks for
  large-scale image recognition,'' \emph{arXiv preprint arXiv:1409.1556}, 2014.

\bibitem{szegedy2015googlenet}
C.~{Szegedy}, {Wei Liu}, {Yangqing Jia}, P.~{Sermanet}, S.~{Reed},
  D.~{Anguelov}, D.~{Erhan}, V.~{Vanhoucke}, and A.~{Rabinovich}, ``Going
  deeper with convolutions,'' in \emph{2015 IEEE Conference on Computer Vision
  and Pattern Recognition (CVPR)}, 2015, pp. 1--9.

\end{thebibliography}

\end{document}